\documentclass[10pt,conference,final]{IEEEtran}
\usepackage{mathrsfs}
\usepackage{bbding}
\usepackage{graphicx}
\usepackage{makecell}
\usepackage{float}
\usepackage{booktabs}
\usepackage[hidelinks]{hyperref}
\usepackage{cite}

\usepackage{enumitem}
\usepackage{color}
\usepackage{ulem}
\usepackage{psfrag}
\usepackage{subfigure}
\usepackage{amssymb}
\usepackage{amsthm}
\usepackage{setspace}
\usepackage{epsfig}
\usepackage{pifont}
\usepackage{amsmath}
\usepackage{array}
\newcommand{\PreserveBackslash}[1]{\let\temp=\\#1\let\\=\temp}
\newcolumntype{C}[1]{$>${\PreserveBackslash\centering}p{#1}}
\newcolumntype{R}[1]{$>${\PreserveBackslash\raggedleft}p{#1}}
\newcolumntype{L}[1]{$>${\PreserveBackslash\raggedright}p{#1}}
\usepackage{multicol}
\usepackage{multirow}
\usepackage{diagbox}
\usepackage{pifont}
\usepackage{indentfirst}
\usepackage{amsfonts}
\usepackage{algorithm}
\usepackage{algorithmicx}
\usepackage{algpseudocode}
\floatname{algorithm}{Procedure}
\usepackage{fancyhdr}
\usepackage{amscd}
\usepackage{xcolor}
\usepackage{bm}
\usepackage{fancyhdr}
\setlist{itemsep=0pt,parsep=0pt}
\allowdisplaybreaks[4]

\makeatletter
\def\endthebibliography{%
	\def\@noitemerr{\@latex@warning{Empty `thebibliography' environment}}%
	\endlist
}
\makeatother
\IEEEoverridecommandlockouts

\usepackage{caption}
\begin{document}
	\title{\huge Cross-System Neural Precoder: \\
		Exploiting Structural Consistency for Fast Adaptation\vspace{-3mm}}
%	\title{\huge Algorithm-Inspired Deep Neural Network for Precoding: \\
%		Are Large Models Necessary for Cross-System Adaptability?\vspace{-3mm}}	
	% PASS, flexible antenna system, scalability, flexiblity
	
	% optimization,
	
	\author{
		% \thanks{This work is supported by National Natural Science Foundation of China (NSFC) under Grant XXX}
		% \IEEEauthorblockN{}.
		%		\thanks{
		%		}
		%	
		\IEEEauthorblockN{Jia Guo and Chenyang Yang}
		
		\IEEEauthorblockA{Beihang University, Beijing, China\\ \{guojia, cyyang\}@buaa.edu.cn \vspace{-3mm}}
	}
	\maketitle
	\setcounter{page}{1}
	\thispagestyle{empty}
	
\begin{abstract}
Adapting learning-based precoding across different system configurations is challenging due to multiple types of variables and constraints. While large-scale neural networks have been proposed for cross-task adaptation, whether such adaptability requires large models remains unclear.
In this paper, we identify a structural property of a class of precoding problems: the subproblems associated with each type of variable in alternative optimization (AO) share a common computational structure across systems when other variables are fixed. This structural consistency enables the reuse of update rules across systems.
Based on this observation, we propose a cross-system neural precoder (XNP), where each layer implements AO-inspired update equations, which define the layer-wise input-output mappings. By reusing common update structures and learning only lightweight nonlinear mappings, the XNP enables efficient adaptation across systems only with several thousand trainable parameters.
Simulation results show that pre-trained XNPs achieve fast adaptation to new configurations with significantly fewer training samples and epochs than a graph neural network-based baseline. This demonstrates that cross-system adaptability can be achieved by exploiting shared computational structure, rather than relying on large models.

%the effectiveness of structural reuse over purely empirical model scaling.
		
		\begin{IEEEkeywords}
			Precoding, cross-system adaptation, structural learning, multi-user multi-antenna systems
		\end{IEEEkeywords}
	\end{abstract}
\section{Introduction}\label{sec:intro}
Precoding is a key technique for enhancing spectral efficiency (SE) in multi-antenna systems. Its design depends on diverse system configurations, including different antenna architectures (e.g., fully digital or hybrid precoding, with or without reconfigurable intelligent surfaces (RIS)) and different antenna settings (e.g., multi-user multi-input multi-output (MU-MIMO) and multi-user multi-input single-output (MU-MISO)). These configurations involve heterogeneous optimization variables, such as digital precoders, analog precoders, and RIS reflection coefficients, under different constraints.

Such diversity makes it challenging to develop learning-based precoding methods adaptable across systems. Existing methods are often tailored to specific configurations.
Recent works have explored large-scale neural networks for adaptation across wireless tasks \cite{wirelessgpt,alikhani2024large}. Many of them adopt architectures from other domains, such as large language models \cite{LLM-LEO} and large vision models \cite{LVM4CSI}, which carry structural priors from their original applications. For example, Transformer-based models are designed for sequential data and have been applied to prediction-oriented wireless tasks \cite{sheng2025wireless}. However, precoding does not naturally exhibit a sequential structure. Whether such architectures are suitable remains unclear.

Another line of work studies foundation models for wireless communications \cite{wirelessgpt,alikhani2024large}. While these approaches demonstrate promising empirical performance, they typically require substantial training data (ranging from millions to hundreds of billions of samples) and computational resources \cite{jiao20246g,wirelessgpt,alikhani2024large}. Moreover, they do not explain
why different systems can share a common architecture.

Model-driven approaches can improve learning efficiency by incorporating domain knowledge into neural networks \cite{DeepUnfold_WMMSE_TWC_2021}. While effective for specific problems, such approaches are usually closely tied to particular formulations. Their adaptability across systems with different variables or
constraints is limited. 
This raises a natural question: \emph{Is there a common structure underlying precoding problems across systems that can be exploited to facilitate cross-system adaptation?}

In this paper, we answer this question affirmatively. Instead of starting from a large generic model, we examine the update steps in alternative optimization (AO), which is widely used for problems involving multiple types of variables. For a class of precoding problems, we find that when all but one type of variables are fixed, the resulting subproblems exhibit a shared computational structure across different system configurations. In particular, although the specific expressions depend on the system model, the objective functions and the corresponding update steps can be expressed in similar forms.  This suggests that the update rules associated with each type of variable can be designed in a reusable manner, which provides the basis for constructing a neural architecture with shared update modules across systems.
Inspired by this observation,
we propose a cross-system neural precoder (XNP), where each
layer corresponds to one round of AO updates. 
To reduce training complexity, the design retains mathematical operations such as matrix multiplications, which are hard to learn but easy to compute. In this way, the XNP focuses on learning lightweight nonlinear transformations  to enhance adaptability. The proposed XNP differs from existing
model-driven approaches. It is built upon cross-system 
structural consistency, and is not tied to a single system configuration within the considered class of precoding problems.
The main contributions are as follows.
\begin{itemize}
\item \textbf{Structure-oriented analysis:} We analyze AO-based precoding algorithms and find that the associated subproblems exhibit similar computational patterns across a class of system configurations when other optimization variables are fixed.

\item \textbf{Reusable architecture:} We develop the XNP, where AO-inspired update rules are implemented as shared neural modules, and variations in variable dimensions and constraint types are handled in a consistent manner. This enables adapting the same architecture across different systems without redesigning the architecture template.

\item \textbf{Efficient adaptation:} We demonstrate that pre-trained XNPs can be adapted to new system configurations with significantly fewer training samples and epochs than an existing neural network. The results indicate that leveraging structural reuse can improve adaptation efficiency beyond standard pre-training alone.
\end{itemize}

	\section{Precoding Problem Formulation}\label{sec:problem}
We optimize precoding in	
multi-antenna systems under different system configurations: fully digital precoding (FP) and hybrid precoding (HP) in fixed-antenna systems (FAS) and RIS-aided systems (RIS system for short). In FAS, there are $K$ users, each equipped with $N_{\sf U}$ antennas, and the base station (BS) equipped with $N_{\sf B}$ antennas transmits $M$ data streams to each user. In RIS system, there is a RIS with $N$ reflective elements, which helps the BS to serve the $K$ users. We take the problem of maximizing the SE as an example.
	
Denote $\mathcal{X}$ as the set of optimization variables. Depending on the system configurations, $\mathcal{X}$ may include digital precoding matrix $\mathbf{V}=[\mathbf{V}_1,\ldots,\mathbf{V}_K]$, the analog precoding matrix $\mathbf{V}_{\rm RF}$, and the RIS reflective coefficient matrix
$
\bm{\Theta}
\triangleq
\mathrm{diag}(\theta_1,\ldots,\theta_N)
$.  The precoding optimization problem in these systems can be expressed in a common formulation as,
	\begin{subequations}\label{eq:sec2_prob_general}
		\begin{align}
			\max_{\mathcal{X}} \quad
			& \sum_{k=1}^{K} R_k(\mathcal{X})
			\label{eq:sec2_prob_general_obj}\\
			\textrm{s.t.} \quad
			& \sum_{k=1}^{K}\|\tilde{\mathbf{V}}_k\|_{\sf F}^2 \le P_{\max},
			\label{eq:sec2_prob_general_cons_v}\\
%			& \sum_{k=1}^{K}\|\mathbf{V}_{\rm RF}\mathbf{V}_k\|_F^2 \le P_{\max},
%			\quad \text{if required},
%			\label{eq:sec2_prob_general_cons_vrf}\\
			& |(\mathbf{V}_{\rm RF})_{i,j}| = 1,\ \forall i,j,
			\quad \text{if required},
			\label{eq:sec2_prob_general_cons_cm_vrf}\\
			& |\theta_n| = 1,\ \forall n,
			\quad \text{if required},
			\label{eq:sec2_prob_general_cons_cm_theta}
		\end{align}
	\end{subequations}
	where $\tilde{\mathbf{V}}_k$ is the equivalent precoding matrix of the $k$-th user, $P_{\max}$ is the maximal transmit power, $(\mathbf{V}_{\rm RF})_{i,j}$ is the element on the $i$-th row and $j$-th column of the analog precoding matrix $\mathbf{V}_{\rm RF}$, $|\cdot|$ denotes the magnitude of a complex value, $(\cdot)^{\sf H}$ and $\|\cdot\|_{\sf F}$ denote the Hermitian transpose and Frobenius norm, respectively, $R_k(\mathcal{X})$ is
the achievable rate of user $k$ that can be expressed as,
	\begin{align}
		&R_k(\mathcal{X})
		=\notag\\
		& \log_2 \det \Bigg(
		\mathbf{I}
		+
		\mathbf{Q}_{k,k}\mathbf{Q}_{k,k}^{\sf H}
		\Big(
		\sum_{j=1,j\neq k}^{K}
		\mathbf{Q}_{k,j}\mathbf{Q}_{k,j}^{\sf H}
		+
		\sigma_0^2 \mathbf{I}
		\Big)^{-1}
		\Bigg),
		\notag
	\end{align}
%	where  $\mathbf{Q}_{k,j}$ is the product of precoding matrix of the $j$-th user and the channel matrix of the $k$-th user,
with $\sigma_0^2$ being the noise power.
	The definitions of $\mathcal{X}$, $\tilde{\mathbf{V}}_k$, $\mathbf{Q}_{k,j}$,  and the required constraints in different systems are respectively listed as follows.
	
	\noindent \textit{1) FP in FAS:}
	In this system,
	$
		\mathcal{X} \triangleq \{\mathbf{V}\}, \tilde{\mathbf{V}}_k \triangleq \mathbf{V}_k,
		\mathbf{Q}_{k,j} \triangleq \mathbf{H}_k \mathbf{V}_j,
	$
	where $\mathbf{H}_k$ is the channel matrix from the BS to user $k$. The required constraint is \eqref{eq:sec2_prob_general_cons_v}.
	
	\noindent \textit{2) HP in FAS:}
	$
		\mathcal{X} \triangleq \{\mathbf{V},\mathbf{V}_{\rm RF}\}, \tilde{\mathbf{V}}_k \triangleq \mathbf{V}_{\rm RF}\mathbf{V}_k,
		\mathbf{Q}_{k,j}
		 \triangleq \mathbf{H}_k \mathbf{V}_{\rm RF} \mathbf{V}_j
	$.
	The required constraints are
	\eqref{eq:sec2_prob_general_cons_v} and
	\eqref{eq:sec2_prob_general_cons_cm_vrf}.
	
	\noindent \textit{3) FP in RIS system:}
	In this system,
	$
		\mathcal{X} \triangleq \{\mathbf{V},\bm{\Theta}\}, \tilde{\mathbf{V}}_k \triangleq \mathbf{V}_k,
		\mathbf{Q}_{k,j}
		\triangleq \big(\mathbf{H}_k^{d}
		+ \mathbf{G}\bm{\Theta}\mathbf{H}_k^{r}\big)\mathbf{V}_j
	$,
	where $\mathbf{H}_k^{d}$ is the BS-user channel, $\mathbf{G}$ is the BS-RIS channel, $\mathbf{H}_k^{r}$ is the RIS-user channel.
	The constraints are
	\eqref{eq:sec2_prob_general_cons_v} and
	\eqref{eq:sec2_prob_general_cons_cm_theta}.
	
	\noindent \textit{4) HP in RIS system:}
	$
		\mathcal{X}
		\triangleq \{\mathbf{V},\mathbf{V}_{\rm RF},\bm{\Theta}\}, \tilde{\mathbf{V}}_k \triangleq \mathbf{V}_{\rm RF}\mathbf{V}_k,
		\mathbf{Q}_{k,j}
		\triangleq \big(\mathbf{H}_k^{d}
		+ \mathbf{G}\bm{\Theta}\mathbf{H}_k^{r}\big)
		\mathbf{V}_{\rm RF}\mathbf{V}_j
	$.
	The constraints are
	\eqref{eq:sec2_prob_general_cons_v},
	\eqref{eq:sec2_prob_general_cons_cm_vrf}, and
	\eqref{eq:sec2_prob_general_cons_cm_theta}.

%	We can see that the considered precoding problems under different systems share the same sum-rate maximization form in \eqref{eq:sec2_prob_general}. The differences lie in the optimization variable set $\mathcal{X}$, $\mathbf{Q}_{k,j}$, and the required constraints.
While the considered precoding problems can be described under a common formulation, the precoding in different systems has the following differences:
\begin{itemize}
    \item {variable set} (e.g., $\mathbf{\Theta}$ only in RIS systems),
    \item {variable dimensions} (e.g., size of $\mathbf{V}$, $\mathbf{V}_{\mathrm{RF}}$, $\mathbf{\Theta}$),
    \item {constraint types} (power constraints appear in all systems, while constant-modulus constraints only exist for analog precoding and RIS reflective coefficients).
\end{itemize}
Nonetheless, such a problem formulation is the starting point for finding the structural consistency across systems.
	
	\section{Shared Computational Structure of AO Subproblems Across Systems}

In this section, we examine the structure of the subproblems arising in alternative optimization for the precoding problems, where each variable is updated while fixing the others.
Since the power and constant-modulus constraints can be satisfied by normalization or projection operations, we focus on analyzing the objective functions of the subproblems, and finding the structure of the iteration equation for each type of variable.

	\subsection{Structure of the Update for Digital Precoders}
	
	By fixing $\mathbf{V}_{\rm RF}$ and $\bm{\Theta}$ (if they need to be optimized), the objective function of the subproblem for optimizing $\mathbf{V}$ can be expressed in a common form as,
	\begin{align}
		\mathcal{O}(\mathbf{V})& \triangleq
		\sum_{k=1}^{K}\log_2\det\!\Bigg(
		\mathbf{I}+\tilde{\mathbf{H}}_k\mathbf{V}_k\mathbf{V}_k^{\sf H}\tilde{\mathbf{H}}_k^{\sf H} \notag\\
		&\Big(\sum_{j=1,j\neq k}^{K}\tilde{\mathbf{H}}_k\mathbf{V}_j\mathbf{V}_j^{\sf H}\tilde{\mathbf{H}}_k^{\sf H}+\sigma_0^2\mathbf{I}\Big)^{-1}
		\Bigg), \label{eq:sec3_v_obj}
	\end{align}
	where $\tilde{\mathbf{H}}_k$ denotes the equivalent channel seen by the digital precoder for the $k$-th user. Specifically,
	\begin{itemize}
		\item For FP in FAS, $\tilde{\mathbf{H}}_k=\mathbf{H}_k$;
		\item For HP in FAS, $\tilde{\mathbf{H}}_k=\mathbf{H}_k\mathbf{V}_{\rm RF}$;
		\item For FP in RIS system, $\tilde{\mathbf{H}}_k=\mathbf{H}_k^{d}+\mathbf{G}\bm{\Theta}\mathbf{H}_k^{r}$;
		\item For HP in RIS system, $\tilde{\mathbf{H}}_k=(\mathbf{H}_k^{d}+\mathbf{G}\bm{\Theta}\mathbf{H}_k^{r})\mathbf{V}_{\rm RF}$.
	\end{itemize}

	Since $\mathcal{O}(\mathbf{V})$ follows a similar computational structure across systems, the equation for finding $\mathbf{V}$ iteratively can be written in a shared form. We next provide an example of the iterative equation.
	
To deal with the non-convex objective function in \eqref{eq:sec3_v_obj}, one way is to minimize the weighted mean-square-error (MSE) between the recovered signal and the original signal as in \cite{WMMSE2011Shi},
	\begin{align}
		{\sf MSE}(\mathbf{W,U,V})=\displaystyle\sum_{k=1}^K {\rm Tr}(\mathbf{W}_k\mathbf{E}_k) - \log\det(\mathbf{W}_k),\label{eq:sec3_mse}
	\end{align}
	where $\mathbf{W}_k$ is the weight matrix, $\mathbf{U}_k$ is the receive combining matrix, and $\mathbf{E}_k$ is the MSE matrix of the $k$-th user.
	
	For notational simplicity, we adopt the approximation in \cite{GJ-MLSP} that
%the inter-stream interference is weak and set
$\mathbf{W}_k\approx\mathbf{I}$.
 %(this assumption will be removed in the simulations).
 Then, only $\mathbf{U}_k$ and $\mathbf{V}_k$ need to be updated.
	The closed-form expressions of $\mathbf{U}_k$ and $\mathbf{V}_k$ can be found by setting the first-order derivative of \eqref{eq:sec3_mse} to zero, which however consists of computationally intensive matrix inversion. To avoid this, we instead use gradient descent to update the variables as,
	\begin{align}
		\mathbf{V}_k \leftarrow\
		& \mathbf{V}_k
		-\lambda \sum_{j=1}^{K}\tilde{\mathbf{H}}_j^{\sf H}\mathbf{U}_j^{\sf H}
		(\mathbf{U}_j\tilde{\mathbf{H}}_j\mathbf{V}_k)
		+\lambda \tilde{\mathbf{H}}_k^{\sf H}\mathbf{U}_k^{\sf H}, \label{eq:sec3_v_update}\\
		\mathbf{U}_k \leftarrow\
		& (1-\lambda\sigma_0^2)\mathbf{U}_k
		-\lambda \sum_{j=1}^{K}
		(\mathbf{U}_j\tilde{\mathbf{H}}_j\mathbf{V}_k)\mathbf{V}_k^{\sf H}\tilde{\mathbf{H}}_j^{\sf H}
		+\lambda \mathbf{V}_k^{\sf H}\tilde{\mathbf{H}}_k^{\sf H}, \label{eq:sec3_u_update}
	\end{align}
	where $\lambda$ is the step size. Here and after, we use ``$\leftarrow$'' instead of ``$=$'' because the meaning of variables in the left- and right-hand sides differs. Taking $\mathbf{V}_k$ as an example, the $\mathbf{V}_k$ on the right-hand side denotes the value in the previous iteration, while the $\mathbf{V}_k$ on the left-hand side denotes the value in the current iteration.
%  to avoid misunderstanding that the variables in the left- and right-hand side of "$\leftarrow$" are the same.
	
	% These equations have the same form across systems once $\tilde{\mathbf{H}}_k$ is specified.
	
	\subsection{Structure of the Update for Analog Precoders}
	
	By fixing $\bm{\Theta}$ (if it needs to be optimized) and $\mathbf{V}$, the subproblem for optimizing $\mathbf{V}_{\rm RF}$ in systems with hybrid precoding can be written as,
	\begin{align}
		\mathcal{O}(\mathbf{V}_{\rm RF})& \triangleq
		\sum_{k=1}^{K}\log_2\det\!\Bigg(
		\mathbf{I}+\tilde{\mathbf{H}}_k\mathbf{V}_{\rm RF}\mathbf{V}_k\mathbf{V}_k^{\sf H}\mathbf{V}_{\rm RF}^{\sf H}\tilde{\mathbf{H}}_k^{\sf H}
		\Big( \notag\\
		&\sum_{j=1,j\neq k}^{K}\tilde{\mathbf{H}}_k\mathbf{V}_{\rm RF}\mathbf{V}_j\mathbf{V}_j^{\sf H}\mathbf{V}_{\rm RF}^{\sf H}\tilde{\mathbf{H}}_k^{\sf H}+\sigma_0^2\mathbf{I}\Big)^{-1}
		\Bigg), \label{eq:sec3_vrf_obj}
	\end{align}
	where
	\begin{itemize}
		\item For HP in FAS, $\tilde{\mathbf{H}}_k=\mathbf{H}_k$;
		\item For HP in RIS systems, $\tilde{\mathbf{H}}_k=\mathbf{H}_k^{d}+\mathbf{G}\bm{\Theta}\mathbf{H}_k^{r}$.
	\end{itemize}
	
%	The same objective function across systems induces the same equation for updating $\mathbf{V}_{\rm RF}$ among all the systems involving analog precoding.
With manifold optimization, the iterative equation  to maximize $\mathcal{O}(\mathbf{V}_{\rm RF})$ in the two systems can be expressed as \cite{GJ-RGNN},
	{\begingroup
	\thinmuskip=0mu
	\medmuskip=0mu
	\thickmuskip=0mu
	\begin{align}
		&\mathbf{V}_{\rm RF} \leftarrow \mathcal{P}_{\rm RF}\Bigg(\!\mathbf{V}_{\rm RF}, \sum_{j=1}^K \tilde{\mathbf{H}}_j^{\sf H}\mathbf{U}_j^{\sf H}\mathbf{V}_j^{\sf H} +\!
		\sum_{j,k=1}^K \tilde{\mathbf{H}}_j^{\sf H}\mathbf{U}_j^{\sf H}(\mathbf{U}_j\tilde{\mathbf{H}}_j\mathbf{V}_k)\mathbf{V}_k^{\sf H} \notag\\
		&\hspace{10mm}+ \frac{\sigma_0^2}{P_{\max}}\sum_{j,k=1}^K \mathbf{V}_{\rm RF}\mathbf{V}_k\mathbf{U}_j\mathbf{U}_j^{\sf H}\mathbf{V}_k^{\sf H}\Bigg), \label{eq:sec3_vrf_update}
	\end{align}\endgroup}
	\noindent where $\mathcal{P}_{\rm RF}(\cdot)$ denotes the projection operation for satisfying the constant-modulus constraint on $\mathbf{V}_{\rm RF}$.
	
	\subsection{Structure of the Update for RIS Coefficients}
	
	By fixing $\mathbf{V}$ and $\mathbf{V}_{\rm RF}$ (if they need to be optimized), the subproblem for optimizing the RIS reflective coefficient matrix $\bm{\Theta}$ can be written as
	\begin{align}
		\mathcal{O}(\bm{\Theta})& \triangleq
		\sum_{k=1}^{K}\log_2\det\!\Bigg(
		\mathbf{I}+\tilde{\mathbf{H}}_k\tilde{\mathbf{V}}_k\tilde{\mathbf{V}}_k^{\sf H}\tilde{\mathbf{H}}_k^{\sf H}
		\Big(\notag\\
		&\sum_{j=1,j\neq k}^{K}\tilde{\mathbf{H}}_k\tilde{\mathbf{V}}_j\tilde{\mathbf{V}}_j^{\sf H}\tilde{\mathbf{H}}_k^{\sf H}+\sigma_0^2\mathbf{I}\Big)^{-1}
		\Bigg), \label{eq:sec3_theta_obj}
	\end{align}
	where $\tilde{\mathbf{H}}_k=\mathbf{H}^{\sf d}_k+\mathbf{G}\bm\Theta\mathbf{H}_k^{\sf r}$.
	\begin{itemize}
		\item For FP in RIS-aided systems, $\tilde{\mathbf{V}}_k=\mathbf{V}_k$;
		\item For HP in RIS-aided systems, $\tilde{\mathbf{V}}_k=\mathbf{V}_{\rm RF}\mathbf{V}_k$.
	\end{itemize}
	
	Again, owing to the common structure of the objective function, the common iterative equation for finding $\bm\Theta$ can be used among the RIS systems, e.g., the equation in \cite{ris-dl},
	\begin{equation}
		\theta_n \leftarrow
		\mathcal{P}_{\theta}\!\left(
		\theta_n,\
		\sum_{k=1}^{K}\sum_{j=1}^{K}
		\mathbf{g}_n^{\sf H}\mathbf{U}_j^{\sf H}
		\big(\mathbf{U}_j\tilde{\mathbf{H}}_j\tilde{\mathbf{V}}_k-\delta_{jk}\mathbf{I}\big)
		\tilde{\mathbf{V}}_k^{\sf H}(\mathbf{h}_{nk}^{r})^*
		\right), \label{eq:sec3_theta_update}
	\end{equation}
	where $\mathcal{P}_{\theta}(\cdot)$ denotes the projection operation for satisfying $|\theta_n|=1$, $\mathbf{g}_n$ is the channel vector from the BS to the $n$-th RIS element, and $\mathbf{h}_{nk}^{r}$ is the channel vector from the $n$-th RIS element to the $k$-th user, $\delta_{jk}=1$ if $j=k$ and $0$ otherwise, $(\cdot)^*$ denotes complex conjugation.

{\bf Remark:}	These analyses show that the objective functions of the AO subproblems and the iterative equations for optimizing $\mathbf{V}$, $\mathbf{V}_{\rm RF}$, and $\bm{\Theta}$ exhibit similar computational structures across systems.
	%	The iterative equations are also the same across different antenna settings, including MU-MIMO, MU-MISO, SU-MIMO. This is because the sub-problems are the same for different values of $M,N,K$, and only the sizes of matrices $\mathbf{V},\mathbf{V}_{\rm RF},\mathbf{U},\bm\Theta$ are different. Hence, the iterative equations in \eqref{eq:sec3_v_update},\eqref{eq:sec3_u_update},\eqref{eq:sec3_vrf_update} and \eqref{eq:sec3_theta_update} are also with the same form for different values of $M,N,K$, only the matrices are with different sizes.
		This provides the basis for designing a neural architecture with reusable update equations across systems.
%, where each layer mimics one round of AO updates and each update equation is associated with one type of variable.

%We show that the subproblems for updating each type of variable share a common structure across systems, after fixing the other variables. %Specifically, the objective functions of these subproblems can be written using the same functional form, while differences in variable dimensions and constraints appear only as differences in matrix sizes and the presence or absence of projection operations.
%This structural consistency provides a principled basis for designing reusable neural update modules.

	\section{XNP: Cross-System Adaptable Neural Network Design}
	
	% In Section III, we have shown that the sub-problems for optimizing different variables can be expressed in unified forms across systems. As a result, the corresponding iterative equations share the same structure, although the dimensions of matrices and the involved variables may differ.
	
%	Based on the observation that the sub-problems for optimizing different variables can be expressed in the same forms across systems,
Inspired by the structural analysis of the iterative equations across systems,
we design the XNP in this section, whose architecture follows the AO updates. Recall that the precoding in different systems has three
differences, but they do not break the structural reusability of the update equations since system-specific features can be absorbed into dimension-adaptive fully-connected neural networks (FNNs) and conditional projection layers. The basic idea of designing the XNP is as follows.
\begin{itemize}
    \item {Variable set difference} can be handled by activating only the relevant update equations in the XNP architecture (e.g., the update of $\mathbf{\Theta}$  is disabled in FAS systems).
    \item {Variable dimension difference} can be handled by the element-wise FNN design, which operates on individual entries of the matrices and adapts to arbitrary input sizes.
    \item {Constraint differences} are handled by conditional activation of projection operations ($\mathcal{P}_{\mathrm{RF}}$, $\mathcal{P}_{\theta}$) after each update. In systems without constant-modulus constraints, these projections are simply omitted.
\end{itemize}
	
	\subsection{Overall Architecture of XNP}
	The XNP contains $L$ layers, where each layer updates different types of variables sequentially. For easy exposition, we denote the update equation for a variable $\mathbf{X}$ in the $\ell$-th layer as $\mathcal{M}_{\mathbf{X}}^{(\ell)}$.
	
	The updated digital precoding matrix, combining matrix, analog precoding matrix, and passive precoding matrix (i.e., RIS reflection coefficient matrix) in the $\ell$-th layer are respectively denoted as $\mathbf{V}^{(\ell)}=[\mathbf{V}_1^{(\ell)},\cdots,\mathbf{V}_K^{(\ell)}], \mathbf{U}^{(\ell)}=[\mathbf{U}_1^{(\ell)},\cdots,\mathbf{U}_K^{(\ell)}], \mathbf{V}_{\rm RF}^{(\ell)}$, and $\mathbf{\Theta}^{(\ell)}$. For learning precoding in a system, all or only a part of them are updated, depending on the variable set $\mathcal{X}$. Taking HP in an RIS system as an example, all of these variables are updated in the $(\ell+1)$-th layer as,
	\begin{align}
		\mathbf{U}^{(\ell+1)} &= \mathcal{M}_{\mathbf{U}}^{(\ell)}(\cdot), \label{eq:upd-u}\\
		\mathbf{V}^{(\ell+1)} &= \mathcal{M}_{\mathbf{V}}^{(\ell)}(\cdot), \label{eq:upd-v}\\
		\mathbf{V}_{\rm RF}^{(\ell+1)} &= \mathcal{M}_{\mathbf{V}_{\rm RF}}^{(\ell)}(\cdot), \\
		\mathbf{\Theta}^{(\ell+1)} &= \mathcal{M}_{\mathbf{\Theta}}^{(\ell)}(\cdot).
	\end{align}
	For learning precoding in other systems, only some of the update equations are ``activated''. For example, only \eqref{eq:upd-u} and \eqref{eq:upd-v} are used for learning FP in FAS.
	
\begin{figure}[!htb]
		\centering
		\includegraphics[width=0.9\linewidth]{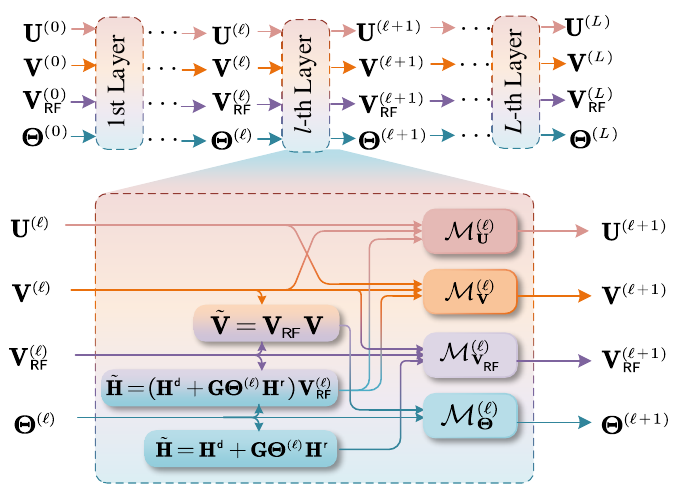}
		\caption{Architecture of XNP.}
		\label{fig:magnet}
	\end{figure}

	Still taking learning HP in the RIS system as an example, the architecture of the XNP is shown in Fig. \ref{fig:magnet}. In each layer of the XNP, the equivalent channel $\tilde{\mathbf{H}}_k$ and precoding matrices are first computed based on the variables in the previous layer, then the variables are updated with the update equations.
	
	% where only the update modules corresponding to the variables in the target system are activated. Therefore, different systems correspond to different combinations of the same set of update modules.
	
	% This architecture explicitly embeds the AO structure into the DNN, instead of learning a direct mapping from channels to precoders. As a result, the inductive bias of the optimization algorithm is preserved.
	\vspace{-2mm}
	\subsection{Design of Update Equations}
	\vspace{-1mm}
	\subsubsection{A straightforward design}
	We take $\mathcal{M}_{\mathbf{V}}^{(\ell)}(\cdot)$ as an example.
	From \eqref{eq:sec3_v_update}, $\mathcal{M}_{\mathbf{V}}^{(\ell)}$ should be a function of $\tilde{\mathbf{H}}_j^{(\ell)},\mathbf{U}_j^{(\ell)},j=1,\cdots,K$ and $\mathbf{V}_k^{(\ell)}$, where $\tilde{\mathbf{H}}_j^{(\ell)}$ is the equivalent channel for the $j$-th user in the $\ell$-th layer.
	
	A straightforward way of designing $\mathcal{M}_{\mathbf{V}}^{(\ell)}$ is to use a neural network, say FNN, to learn the function, i.e.,
	\begin{align}
		\mathcal{M}_{\mathbf{V}}^{(\ell)}: \mathbf{V}_k^{(\ell+1)} = {\sf FNN}(\mathbf{V}_k^{(\ell)}, \tilde{\mathbf{H}}_j^{(\ell)},\mathbf{U}_j^{(\ell)}|j=1,\cdots,K),\notag
	\end{align}
	where ${\sf FNN}(\cdot)$ denotes the input-output (I-O) relation of the FNN.
	
	However, the FNN cannot adapt to different MIMO configurations, where the sizes of matrices $\mathbf{V}_k^{(\ell)}, \tilde{\mathbf{H}}_j^{(\ell)},\mathbf{U}_j^{(\ell)}$ differ. This is because the FNN in each update equation cannot adapt to different input sizes.
	
	We can instead design a neural network with parameter sharing to learn the function for adapting to different input sizes. Nonetheless, such a neural network needs a lot of training samples to learn the  matrix multiplication $\mathbf{U}_j\tilde{\mathbf{H}}_j\mathbf{V}_k$ in \eqref{eq:sec3_v_update},\footnote{	Learning matrix multiplication operation requires high training overhead because it is a non-monotonic function that needs to be approximated with multi-layer neural network. For easy exposition, we explain this by taking learning the multiplication of two $1\times 1$ matrices (i.e., scalars) as an example, i.e., $f(x,y)=xy$. For a single-layered FNN with $x$ and $y$ as inputs, the output of it is $\sigma(w_1x+w_2y)$, where $\sigma(\cdot)$ is an activation function and $w_1, w_2$ are trainable parameters. Because commonly-used $\sigma(\cdot)$ does not affect the monotonicity, the FNN is monotonic in $x$. However, $xy$ is a non-monotonic function of $x$ that increases with $x$ when $y>0$ and decreases with $x$ when $y<0$. Such a function has to be approximated by a multi-layer FNN.}
%Increasing the number of layers improves the approximation ability, but suffers from higher training overhead.
%Learning the multiplication of multiple matrices is even harder.}
which reflects inter-data stream interference  \cite{guo2025attention}.
%that is important for learning beamforming \cite{guo2025attention}.

	% The multiplication result reflects inter-data stream interference that is important for learning beamforming \cite{guo2025attention}.

	% From \eqref{eq:sec3_v_update}, the update of $\mathbf{V}$ depends on the structured operations, such as matrix products that can capture multi-user interference.
	
	\subsubsection{A model-aided neural network design}
	
	% To let the DNN adaptable to different input sizes, parameter sharing can be introduced into the FNN \cite{LSJ}, yet learning matrix multiplications is still with high training overhead. Noticing that the matrix multiplication operations can be applied to different input sizes,
	
	To enable fast cross-system adaptation,
	we retain the matrix multiplications and only learn a lightweight nonlinear mapping, instead of learning the entire mapping $\mathcal{M}_{\mathbf{V}}^{(\ell)}(\cdot)$ with a black-box neural network. Specifically, the update equation is designed as,
	% The iterative equations in \eqref{eq:sec3_v_update}, \eqref{eq:sec3_u_update}, \eqref{eq:sec3_vrf_update} and \eqref{eq:sec3_theta_update} can be adapted to different sizes, because the operations in the equations, including multiplications of channel and precoding matrices and the matrix addition, are the same for different matrix sizes. To make the DNN adaptable to different input sizes and to reduce the training overhead for learning matrix multiplications,  we do not learn
	% Hence, instead of learning the entire mapping $\mathcal{M}_{\mathbf{V}}^{(\ell)}(\cdot)$ with a black-box DNN, we retain the matrix multiplications and only learn a lightweight nonlinear mapping on top of them.
	\begin{align}
		&\mathcal{M}_{\mathbf{V}}^{(\ell)}: \mathbf{V}_k^{(\ell+1)} = {\sf FNN}_{\mathbf{V},\mathrm{ele}}^{(\ell)}\Bigg(\mathbf{V}_k^{(\ell)}, \notag\\
		&\displaystyle\sum_{j=1}^K \tilde{\mathbf{H}}_j^{{(\ell)}\sf H}\mathbf{U}_j^{{(\ell)}\sf H}(\mathbf{U}_j^{(\ell)}\tilde{\mathbf{H}}_j^{(\ell)}\mathbf{V}_k^{(\ell)}), \tilde{\mathbf{H}}_k^{{(\ell)}\sf H}\mathbf{U}_k^{{(\ell)}\sf H}\Bigg),\label{eq:module-v}
	\end{align}
	where ${\sf FNN}_{\mathbf{V},\mathrm{ele}}^{(\ell)}(\cdot)$ denotes an element-wise function with the I-O relation of each element being a FNN, and the FNN is identical among elements. For example, for $\mathbf{y}={\sf FNN}_{\mathbf{V},\mathrm{ele}}^{(\ell)}(\mathbf{x})$ where $\mathbf{x}=[x_1,\cdots,x_K]^{\sf T}$ and $\mathbf{y}=[y_1,\cdots,y_K]^{\sf T}$, it can be expressed as
	$\mathbf{y} = [\mathsf{FNN}(x_1), \cdots,\mathsf{FNN}(x_K)]^{\sf T}$.
	
%	\begin{align}
%		\begin{bmatrix}
%		y_1\\ \vdots \\ y_K
%		\end{bmatrix} = \begin{bmatrix}
%		\mathsf{FNN}(x_1)\\ \vdots \\ \mathsf{FNN}(x_K)
%	\end{bmatrix}. \notag
%	\end{align}
	
The element-wise parameterized function ${\sf FNN}_{\mathbf{V},\mathrm{ele}}^{(\ell)}(\cdot)$ learns the element-wise operations in \eqref{eq:sec3_v_update}, including matrix subtraction and multiplying a scalar on a matrix, which are the same for all the elements. The role of the  FNN is to learn a effective update direction and step size.
%	The reason we use an element-wise parameterized function ${\sf FNN}_{\mathbf{V},\mathrm{ele}}^{(\ell)}(\cdot)$ with identical elements is that the FNNs mimic element-wise operations in \eqref{eq:sec3_v_update}, including matrix subtraction and multiplying a scalar on a matrix, and the operations are the same for all the elements.
	% ${\sf FNN}_{\mathbf{V},\mathrm{ele}}^{(\ell)}(\cdot)$ is identical for all the elements, because the operations it mimics are also identical for all the elements.
	Such a design enables the XNP to adapt to different input sizes, and hence adapt to different antenna settings.

	% This design preserves the structured operations in the iterative equations while using the neural network to learn data-driven update corrections, such as effective step sizes. As a result, the learnable part has low complexity and does not depend on the matrix dimensions.
	
	The update equations for $\mathbf{U}$, $\mathbf{V}_{\rm RF}$, and $\mathbf{\Theta}$ are designed in a similar way as follows,
	\begin{align}
		\mathcal{M}_{\mathbf{U}}^{(\ell)}: &\mathbf{U}_k^{(\ell+1)} = {\sf FNN}_{\mathbf{U},\mathrm{ele}}^{(\ell)}\Bigg(\mathbf{U}_k^{(\ell)}, \displaystyle\sum_{j=1}^K(\mathbf{U}_j^{(\ell)}\tilde{\mathbf{H}}_j^{(\ell)}\mathbf{V}_k^{(\ell)})\cdot\notag\\ &\mathbf{V}_k^{{(\ell)}\sf H}\tilde{\mathbf{H}}_j^{{(\ell)}\sf H}, \mathbf{V}_k^{{(\ell)}\sf H}\tilde{\mathbf{H}}_k^{{(\ell)}\sf H}\Bigg). \label{eq:module-u}\\
		\mathcal{M}_{\mathbf{V}_{\rm RF}}^{(\ell)}:& \mathbf{V}_{\rm RF}^{(\ell+1)} = {\sf FNN}_{\mathbf{V}_{\rm RF},\mathrm{ele}}^{(\ell)}\Bigg(\mathbf{V}_{\rm RF}^{(\ell)}, \displaystyle\sum_{j=1}^K \tilde{\mathbf{H}}_j^{{(\ell)}\sf H}\mathbf{U}_j^{{(\ell)}\sf H}\mathbf{V}_j^{{(\ell)}\sf H},\notag\\
		&\displaystyle\sum_{j=1}^K\sum_{k=1}^K \tilde{\mathbf{H}}_j^{(\ell)\sf H}\mathbf{U}_j^{(\ell)\sf H}(\mathbf{U}_j^{(\ell)}\tilde{\mathbf{H}}_j^{(\ell)}\mathbf{V}_k^{(\ell)})\mathbf{V}_k^{{(\ell)}\sf H}, \notag\\
		&\displaystyle \frac{\sigma_0^2}{P_{\max}}\sum_{j=1}^K\sum_{k=1}^K \mathbf{V}_{\rm RF}^{(\ell)}\mathbf{V}_k^{(\ell)}\mathbf{U}_j^{(\ell)}\mathbf{U}_j^{{(\ell)}\sf H}\mathbf{V}_k^{{(\ell)}\sf H}\Bigg), \label{eq:module-vrf}\\
		\mathcal{M}_{\mathbf{\bm \Theta}}^{(\ell)}:& \displaystyle\theta_n^{(\ell+1)} = {\sf FNN}_{\theta,\mathrm{ele}}^{(\ell)}\Bigg(\theta_n^{(\ell)},
		\sum_{k=1}^K\sum_{j=1}^K \mathbf{g}_n^{\sf H}\mathbf{U}_j^{{(\ell)}\sf H}\big(\notag\\
		&\mathbf{U}_j^{(\ell)}\tilde{\mathbf{H}}_j^{(\ell)}\tilde{\mathbf{V}}_k^{(\ell)}-\delta_{jk}\mathbf{I}\big)\tilde{\mathbf{V}}_k^{{(\ell)}\sf H}(\mathbf{h}_{nk}^{\sf r})^*\Bigg).\label{eq:module-theta}
	\end{align}
	
	After each update, normalization is applied to ensure that the power and constant-modulus constraints are satisfied.
	
%	\subsection{Cross-system adaptability}
%	
%	The cross-system adaptability of XNP stems from the fact that the structured terms in the update modules are derived from unified sub-problems, and hence have the same form across systems. The only difference lies in the instantiated matrices (e.g., equivalent channels) and their dimensions.
%	
%	The learnable part of each module is an element-wise mapping applied to these structured terms. Since the same types of structured quantities appear across systems, the learned mappings can be reused across systems. Therefore, XNP transfers not a full end-to-end mapping, but the learned corrections to a common AO update structure.
%	
%	Moreover, since the element-wise neural mappings are independent of matrix sizes, the same modules can be directly applied when the numbers of antennas, users, or RF chains change.
	
	\subsection{Adaptation Across Systems}
	
An XNP trained in a system (called source system) can be adapted to other systems (called target systems). For a target system with some optimization variables being identical in structure (but possibly differing in dimensions) to those in the source system, the corresponding update equations can be reused.
%The element-wise FNN design automatically handles dimensional differences, while constraint differences are absorbed by activating or deactivating the corresponding projection operations.
Only for variables that do not exist in the source system (e.g., $\bm\Theta$
in a RIS target system when the source is an FAS system), the corresponding FNNs need to be trained.
%For the FNNs in the update equations for other variables that are not trained in the source system, they need to be trained for the target system.
	
	For example, an XNP trained for FP in FASs learns $\mathcal{M}_{\mathbf{V}}^{(\ell)}(\cdot)$ and $\mathcal{M}_{\mathbf{U}}^{(\ell)}(\cdot)$. These equations can be reused for FP in RIS systems, while the FNN in the equation for updating $\bm\Theta$ needs to be trained.
	
	Since the FNNs in only a part of update equations need to be trained, the XNP can adapt to a target system rapidly.
% with reduced training overhead, compared with a directly-trained XNP from scratch.
	
	% All the update equations can then be fine-tuned jointly with a small number of samples from the target system. In this way, the XNP can efficiently adapt to new systems with reduced training overhead.
	
	\section{Simulation Results}
	In this section, we evaluate the cross-system adaptability of the proposed XNP.
	
	Consider a multi-user system. The small-scale channels are Rician channels with a factor of 10, which consist of line-of-sight (LoS) and non-LoS components. The path-loss model from the BS to the users is $32.6+36.7\log_{10}(d)$, and the model from the BS to RIS and from the RIS to the users is $30+22\log_{10}(d)$, where $d$ is the distance in meters. $P_{\max}=30$ dBm, $\sigma_0^2=-80$ dBm.
	
	To pre-train the FNNs in the update equations, we train two XNPs with $L$ layers in two source systems,
	\begin{enumerate}
		\item HP in MU-MISO FAS, where $K=3, N_{\sf B}=8$, the number of RF chains of the BS is $N_{\rm RF}=6$. In this XNP, $\mathcal{M}_{\mathbf{V}_{\rm RF}}^{(\ell)}$ and $\mathcal{M}_{\mathbf{V}}^{(\ell)}$ are trained.
%, $*=1,\cdots,L$.
		\item FP in MU-MISO RIS, where $K=3, N_{\sf B}=8$, the number of reflective elements at the RIS is $N = 10$. In this XNP, $\mathcal{M}_{\mathbf{V}}^{(\ell)}$ and $\mathcal{M}_{\mathbf{\bm\Theta}}^{(\ell)}$ are trained.
	\end{enumerate}
	
	To respectively show the adaptability to different antenna architectures and settings, the trained updated equations are reused in the XNPs for two target systems as follows.
	\begin{enumerate}
	\item HP in MU-MISO RIS, where $K=3, N_{\sf B}=8, N_{\rm RF}=6, N = 10$. This system differs from ``HP in MU-MISO FAS'' in antenna architecture.
%From this we can see the adaptability to different antenna architectures.
	\item FP in MU-MIMO FAS, where $K=3, N_{\sf B}=8, N_{\sf U}=2,M=2$. This system differs from ``FP in MU-MISO RIS'' in antenna settings.
	\end{enumerate}

	The XNPs are trained in an unsupervised manner, where the negative SE averaged over all the training samples is used as the loss function.
	The power and constant-modulus constraints can be satisfied by designing activation functions, as detailed in \cite{GJ-RGNN}.
	The XNPs for the source systems are trained with 10,000 samples. Each sample only contains the channel matrices generated from the channel model. For FAS systems, only the BS-user channels are generated. For RIS systems, the BS-RIS, RIS-user, BS-user channels are generated.

The fine-tuned hyper-parameters are provided in Table \ref{tab:hyper-params}, where $J^{(\ell)}$ denotes the input dimension of each element-wise FNN in the $\ell$-th layer. 	With the hyper-parameters, $\mathcal{M}_{\mathbf{V}}^{(\ell)}$, $\mathcal{M}_{\mathbf{V}_{\rm RF}}^{(\ell)}$, $\mathcal{M}_{\bm\Theta}^{(\ell)}$ in the six layers have 2.6K, 2.2K and 2.6K trainable parameters in total, respectively. %The numbers are much lower than that for a large model.
	
	\begin{table}[!htb]
		\centering
		\caption{Hyper-parameters of XNP}\label{tab:hyper-params}
		\footnotesize
		\begin{tabular}{cl|l}
			\hline\hline
			\multicolumn{2}{c|}{$L$}                                                                                                                                                      & \multicolumn{1}{c}{6}                    \\ \hline
			\multicolumn{2}{c|}{Num. of neurons in   hidden layers}                                                                                                                          & \multicolumn{1}{c}{{[}16, 16, 16, 16, 16, 16{]}} \\ \hline
			\multicolumn{1}{c|}{\multirow{3}{*}{\begin{tabular}[c]{@{}c@{}}Num. of neurons in FNN of \\      every update equation\end{tabular}}} & $\mathcal{M}_{\mathbf{V}}^{(\ell)}$          & $[J^{(\ell)}\times 3,   J^{(\ell)}]$      \\ \cline{2-3}
			\multicolumn{1}{c|}{}                                                                                                                 & $\mathcal{M}_{\mathbf{V}_{\rm RF}}^{(\ell)}$ & $[J^{(\ell)}\times 3,   8, J^{(\ell)}]$   \\ \cline{2-3}
			\multicolumn{1}{c|}{}                                                                                                                 & $\mathcal{M}_{\bm\Theta}^{(\ell)}$           & $[J^{(\ell)}\times 3,   J^{(\ell)}]$      \\ \hline\hline
		\end{tabular}
	\end{table}

	We evaluate the performance of XNP on the target systems.
	The learning performance is measured by SE ratio, which is the SE achieved by the XNP divided by the SE achieved by a numerical algorithm. The algorithm is the block gradient descent algorithm proposed in \cite{RIS_BCD} for RIS, and the weighted-minimum-mean-square-error algorithm in  \cite{WMMSE2011Shi} for FAS.
	
To show the benefit of reusing update equations, the performance of the pre-trained XNP (with legend ``XNP, Tune'') is compared with XNPs directly trained under target systems without pre-training (with legend ``XNP, Direct'').

To show the benefit of the AO-based architecture, we compare XNP with the following baselines.
	\begin{itemize}
		\item \emph{GNN, Direct}: The graph neural network (GNN) proposed in \cite{LSJ} is used to learn the precoding policy, which is trained and tested for every system. We adopt this neural network instead of other non-AO-based architectures (say FNN), because it is with lower training complexity.
		\item \emph{GNN, Tune}: After training a GNN in a source system, the learned weight matrices serve as the initial value for fine-tuning in the target systems.
	\end{itemize}

	For fair comparison, the numbers of layers and neurons in each layer of the GNN are the same as those of the XNP.
	
%	\subsection{Adaptability From Source-1 to Target-1}
%	In Fig. \ref{fig:perf-ntr1}, we show how the SE ratio on the test set changes with the number of training samples. The achieved performance after 1, 50, 100 and 500 epochs is provided in the figure. We can see that by pre-training the update equations from the source system, the XNP can achieve better performance than the one without pre-training. This indicates that the pre-trained XNP needs fewer samples or epochs to achieve the same SE ratio. For example, the pre-trained XNP only needs 50 samples and 50 epochs to achieve over 80\% SE ratio, while the XNP without pre-training requires 100 samples and 500 epochs, or 200 samples and 100 epochs to achieve over 80\% SE ratio.
%	
%	\begin{figure}[!htb]
%		\centering
%		\includegraphics[width=.9\linewidth]{fig-h-non-ris.pdf}
%		\caption{Learning performance of XNP for learning HP in MU-MISO FAS system.}
%		\label{fig:perf-ntr1}
%	\end{figure}
	
\begin{figure}[!htb]
		\centering
		\includegraphics[width=\linewidth]{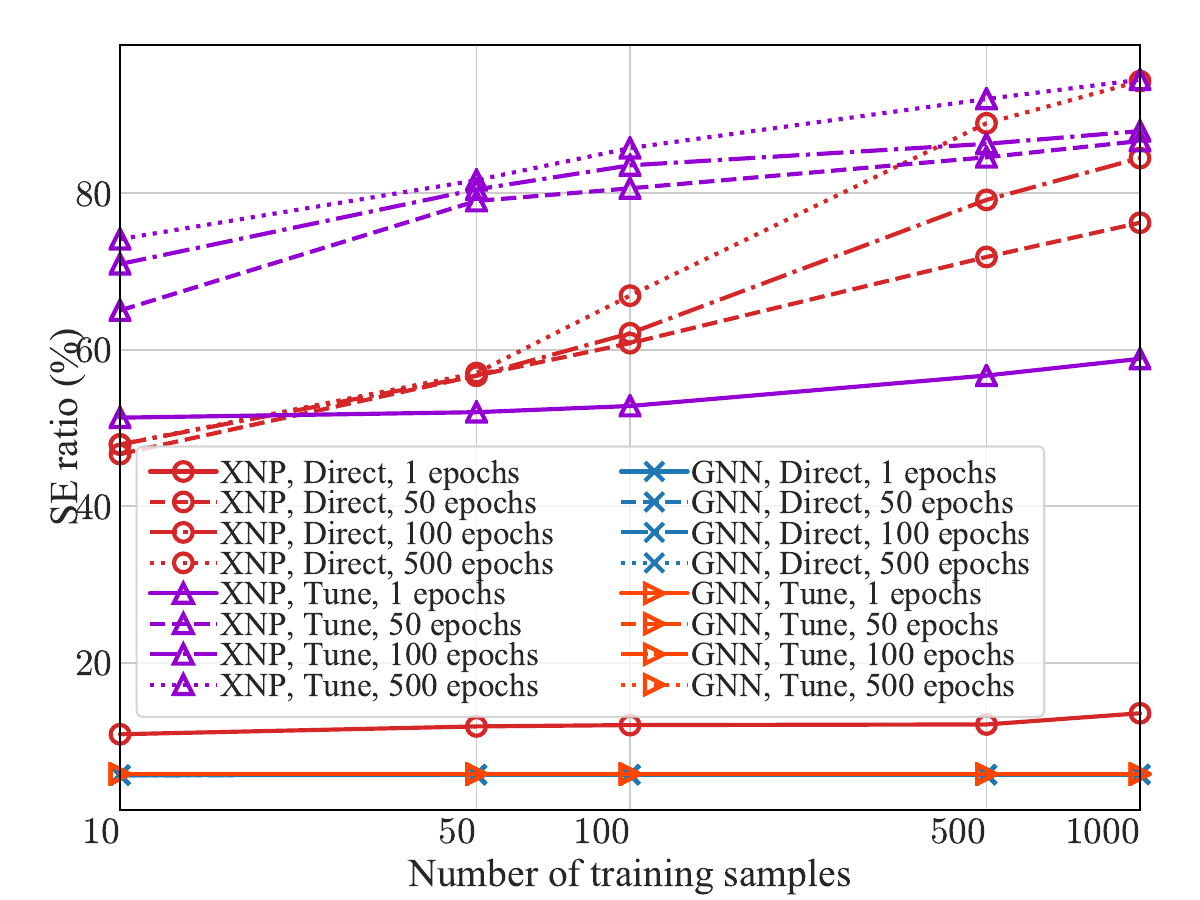}
		\caption{Performance for HP in MU-MISO RIS system. }
		\label{fig:perf-ntr2}
	\end{figure}

	In Fig. \ref{fig:perf-ntr2}, we evaluate the adaptability across antenna architectures, by testing the performance of the HP in MU-MISO RIS system.
%how the SE ratio on the test set changes with the number of training samples.
	The achieved performance after 1, 50, 100 and 500 epochs is provided in the figure.
	% We can see that the pre-training provides good initial parameters for the XNP, such that it can achieve 50\% SE ratio even with only one epoch.
	We can see that the pre-trained XNP can achieve the same SE ratio with fewer samples or epochs than an XNP without pre-training.
	For example, the pre-trained XNP only needs 100 samples and 50 epochs to achieve over 80\% SE ratio, while the XNP without pre-training requires  1000 samples and 100 epochs to achieve over 80\% SE ratio.
	
	It can also be seen that the GNN achieves much lower SE ratio than the XNP, no matter if the GNN is pre-trained or not (all the curves of the GNN are overlapped). This validates the superiority of the AO-based XNP architecture for cross-system adaptability.
	
\begin{figure}[!htb]
		\centering
		\includegraphics[width=\linewidth]{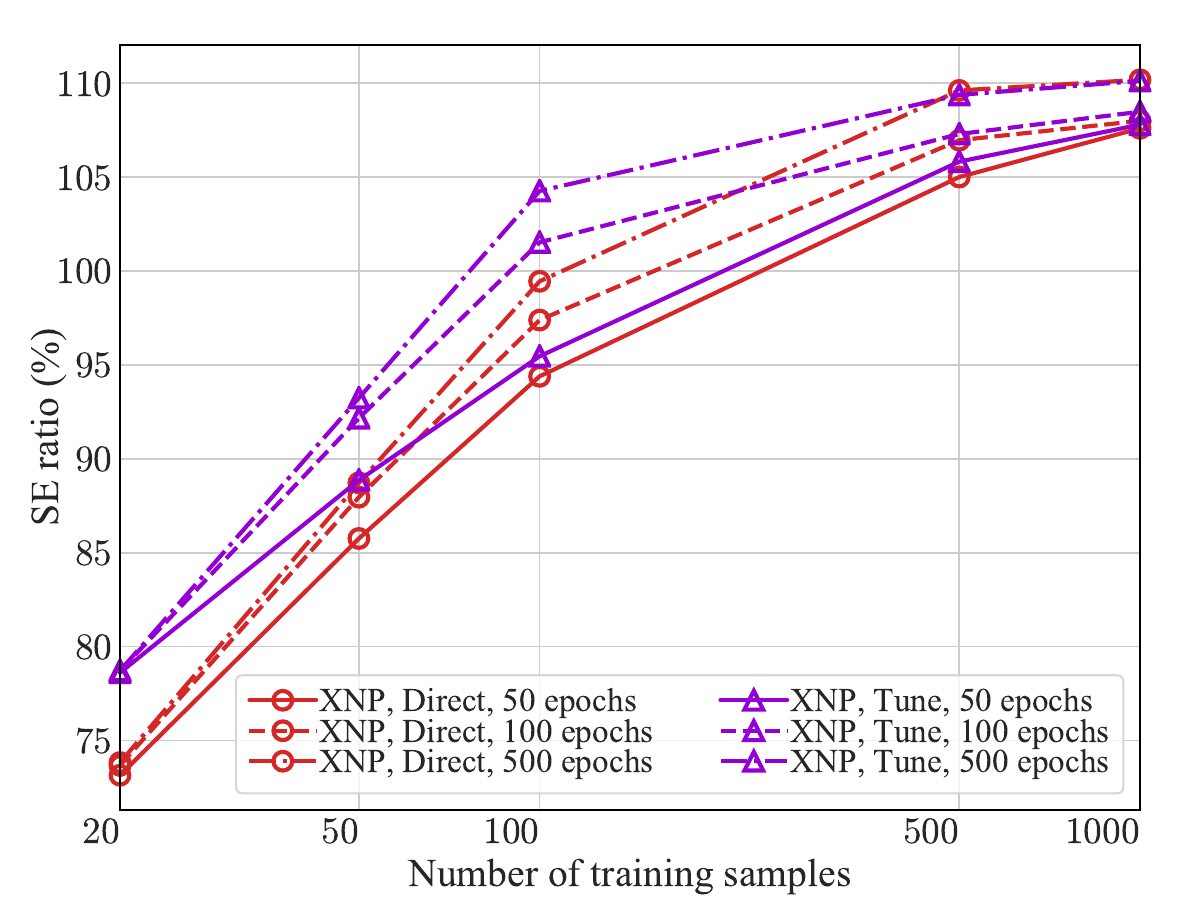}
		\caption{Performance of XNP for FP in MU-MIMO FAS system.}
		\label{fig:perf-ntr3}
	\end{figure}

In Fig. \ref{fig:perf-ntr3}, we evaluate the adaptability across antenna settings, by testing the performance of the FP in MU-MIMO FAS.  It can still be seen that the pre-trained XNP can achieve higher SE than the directly-trained XNP with the same number of training samples and epochs. The SE achieved with only one epoch of the XNP and the SE of the GNN are not provided due to their poor~performance.

\begin{figure}[!htb]
	\centering
	\begin{minipage}[t]{0.48\linewidth}
		\vspace{0pt}
		\centering
		\includegraphics[width=\linewidth]{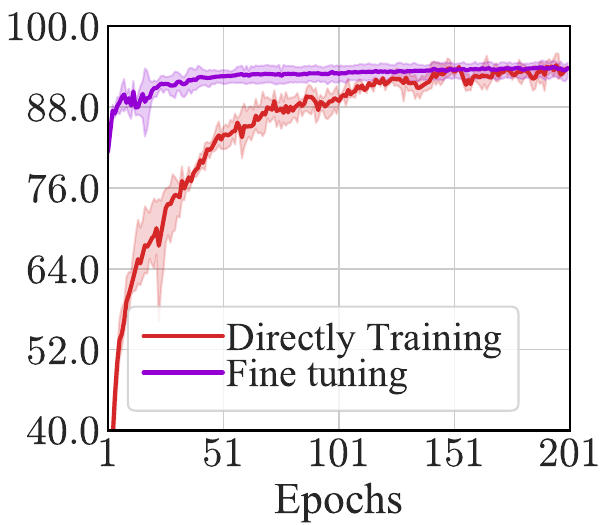}
		\caption{Learning curve when the target system is HP in MU-MISO RIS.}
		\label{fig:perf-epoch}
	\end{minipage}
\hspace{1mm}
	\begin{minipage}[t]{0.48\linewidth}
		\vspace{0pt}
		\centering
		\includegraphics[width=\linewidth]{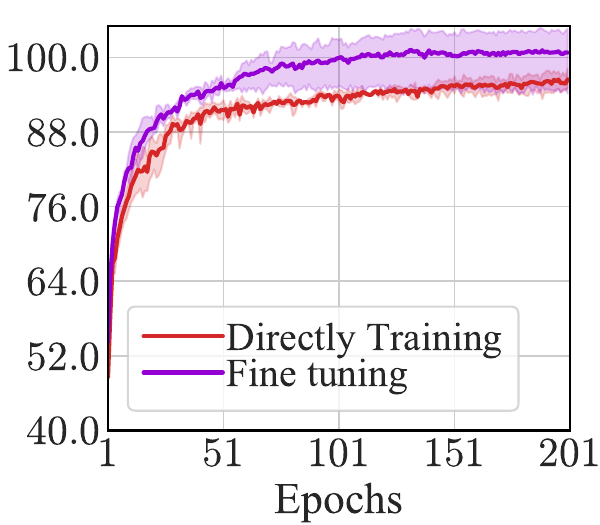}
		\caption{Learning curve when the target system is FP in MU-MIMO FAS.}
		\label{fig:perf-epoch1}
	\end{minipage}
\end{figure}

	In Fig. \ref{fig:perf-epoch}, we provide the learning curves to show the impact of pre-training, where the target system is HP in MU-MISO RIS. For fair comparison, the XNPs are trained with 10,000 samples for both directly training and fine tuning. The curves and the shadowed regions respectively show the mean value and the dynamic region of the achieved SE. The performance of the GNN is not provided, because it does not perform well when trained with these number of samples.
	It can be seen that with pre-training, the XNP only needs less than 10 epochs to achieve 90\% SE ratio. By contrast, the XNP without pre-training needs about 100 epochs to achieve the same performance.

	%	Table \ref{table:hp-ris} shows the training time and number of epochs required for achieving an expected performance (set by 90\% performance ratio). It can be seen that after pre-training from the source systems, the XNP only needs about 30 s and 100 samples to adapt to the target system, while the directly-trained XNP needs at least 193 s and 1,000 samples to learn a good precoding policy. With the growth of the training samples, the expected training time exhibits a decreasing-then-increasing trend. This is because the number of required epochs decreases with the number of training samples, but the time required in every epoch increases.
	%	\begin{table}[!htb]
		%		\setlength\tabcolsep{1pt}
		%		\caption{Time (s) and number of epochs required for adaptation}\label{table:hp-ris}
		%		\footnotesize
		%		\centering
		%		\begin{tabular}{c|c|c|c|c|c|c|c|c|c|c}
			%			\hline\hline
			%			Method & \makecell{Number of\\ training \\samples} & 100 & 200 & 400 & 600 & 800 & 1000 & 2000 & 5000 & 10000\\
			%			\hline
			%			\multirow{2}{*}{\makecell{Fine\\tuning}} & Time & 32.0 & 30.7 & 27.5 & 32.7 & 34.5 & 37.2 & 40.0 & 52.2 & 36.7 \\
			%			\cline{2-11}
			%			& Epochs & 105 & 59 & 36 & 23 & 19 & 15 & 10 & 5 & 1\\
			%			\hline
			%			\multirow{2}{*}{\makecell{Directly\\training}} & Time & $>$300 & $>$300 & $>$300 & $>$300 & 213.2 & 193.5 & 567.6 & 937.4 & 682.4 \\
			%			\cline{2-11}
			%			& Epochs & $>$200 & $>$200 & $>$200 & $>$200 & 140 & 153 & 139 & 94 & 33 \\
			%			\hline\hline
			%		\end{tabular}
		%		``$>$'' indicates that the XNP cannot achieve the expected performance even after the training time and epochs.
		%	\end{table}

	In Fig. \ref{fig:perf-epoch1}, we show the learning curve of the XNP, where the target system is FP in MU-MIMO FAS.  As can be seen from the figure, a pre-trained XNP can achieve 90\% SE ratio with  about 20 epochs, while a directly-trained XNP needs about 40 epochs to achieve the same performance. Moreover, the pre-trained XNP can achieve higher SE than a directly trained one after convergence.
This is probably because that the pre-training helps find a better initial weights to avoid converging to a local minimum point.  These results again
	demonstrate that pre-training the XNPs helps cross-system adaptation.

	% Table \ref{table:hp-ris}
	
	% \begin{table}[!htb]
		% \setlength\tabcolsep{1.5pt}
		% \caption{Time (s) and number of epochs required for adaptation}\label{table:hp-ris}
		% \footnotesize
		% \centering
		% \begin{tabular}{c|c|c|c|c|c|c|c|c|c|c}
			% \hline\hline
			%     Method & \makecell{Number of\\ training \\samples} & 100 & 200 & 400 & 600 & 800 & 1000 & 2000 & 5000 & 10000\\
			%     \hline
			%     \multirow{2}{*}{\makecell{Fine\\tuning}} & Time & 32.0 & 30.7 & 27.5 & 32.7 & 34.5 & 54.2 & 40.0 & 52.2 & 36.7 \\
			%     \cline{2-11}
			%     & Epochs & 105 & 59 & 36 & 23 & 19 & 15 & 10 & 5 & 1\\
			%     \hline
			%     \multirow{2}{*}{\makecell{Directly\\training}} & Time & $>$300 & $>$300 & $>$300 & $>$300 & 213.2 & 193.5 & 567.6 & 937.4 & 682.4 \\
			%     \cline{2-11}
			%     & Epochs & $>$200 & $>$200 & $>$200 & $>$200 & 140 & 153 & 139 & 94 & 33 \\
			%     \hline\hline
			% \end{tabular}
		% \end{table}
	
	\section{Conclusions}
This work studies cross-system adaptation for precoding from a structural viewpoint. Instead of treating different system configurations as independent learning tasks, we examine the update procedures in alternative optimization (AO) and observe that they exhibit similar computational patterns across a class of systems.
Motivated by this observation, we develop a neural architecture, XNP, that constructs its layers according to AO-inspired update steps. Lightweight nonlinear mappings are introduced to enhance flexibility across different variable dimensions and constraints.
Simulation results show that the proposed design can be adapted to new system configurations with reduced training overhead compared with a neural network baseline. These findings suggest that leveraging shared computational structure can be an effective approach to improve adaptation efficiency, offering an alternative to scaling model size.
Overall, the results indicate that incorporating problem structure into learning model design is a promising direction for building efficient and adaptable solutions in wireless communications.

%This work provides a structural perspective on cross-system adaptability for precoding. Rather than treating different system configurations as separate learning tasks, we show that their underlying optimization procedures share common update patterns. 
%Based on this insight, we develop a neural architecture, XNP, which mirrors the iterative optimization process while preserving its low complexity algebraic operations.
%%The resulting design avoids learning fundamental operations and instead focuses on learning lightweight transformations around them.
%This mechanism explains why the XNP can adapt efficiently when system configurations change.
% These results highlight that fast cross-system adaptation can be achieved by exploiting structural invariance in optimization, rather than relying on large-scale neural networks.
%By aligning the network design with the optimization structure, the proposed approach achieves both efficiency and adaptability across systems. This highlights the importance of structure-aware learning model design for wireless problems.

	\begin{spacing}{.9}
		\bibliography{IEEEabrv,GJ}

% Generated by IEEEtran.bst, version: 1.13 (2008/09/30)
\begin{thebibliography}{10}
\providecommand{\url}[1]{#1}
\csname url@samestyle\endcsname
\providecommand{\newblock}{\relax}
\providecommand{\bibinfo}[2]{#2}
\providecommand{\BIBentrySTDinterwordspacing}{\spaceskip=0pt\relax}
\providecommand{\BIBentryALTinterwordstretchfactor}{4}
\providecommand{\BIBentryALTinterwordspacing}{\spaceskip=\fontdimen2\font plus
\BIBentryALTinterwordstretchfactor\fontdimen3\font minus
  \fontdimen4\font\relax}
\providecommand{\BIBforeignlanguage}[2]{{%
\expandafter\ifx\csname l@#1\endcsname\relax
\typeout{** WARNING: IEEEtran.bst: No hyphenation pattern has been}%
\typeout{** loaded for the language `#1'. Using the pattern for}%
\typeout{** the default language instead.}%
\else
\language=\csname l@#1\endcsname
\fi
#2}}
\providecommand{\BIBdecl}{\relax}
\BIBdecl

\bibitem{wirelessgpt}
T.~Yang, P.~Zhang, M.~Zheng \emph{et~al.}, ``{WirelessGPT}: A generative
  pre-trained multi-task learning framework for wireless communication,''
  \emph{IEEE Netw.}, vol.~39, no.~5, pp. 58--65, Sep. 2025.

\bibitem{alikhani2024large}
S.~Alikhani, G.~Charan, and A.~Alkhateeb, ``Large wireless model {(LWM)}: A
  foundation model for wireless channels,'' \emph{arXiv:2411.08872}, 2024.

\bibitem{LLM-LEO}
Z.~Chen, H.~Shin, A.~Nallanathan, and J.~Chambers, ``Large language
  model-empowered channel prediction and predictive beamforming for leo
  satellite communications,'' \emph{arXiv:2510.10561}, 2025.

\bibitem{LVM4CSI}
J.~Guo, P.~Jiang, C.-K. Wen, S.~Jin, and J.~Zhang, ``Lvm4csi: Enabling direct
  application of pre-trained large vision models for wireless channel tasks,''
  \emph{arXiv preprint arXiv:2507.05121}, 2025.

\bibitem{sheng2025wireless}
Y.~Sheng, J.~Wang, X.~Zhou, L.~Liang, H.~Ye, S.~Jin, and G.~Y. Li, ``A wireless
  foundation model for multi-task prediction,'' \emph{arXiv:2507.05938}, 2025.

\bibitem{jiao20246g}
T.~Jiao, C.~Ye, Y.~Huang, Y.~Feng, Z.~Xiao, Y.~Xu, D.~He, Y.~Guan, B.~Yang,
  J.~Chang \emph{et~al.}, ``{6G}-oriented {CSI}-based multi-modal pre-training
  and downstream task adaptation paradigm,'' \emph{IEEE ICC Workshops}, 2024.

\bibitem{DeepUnfold_WMMSE_TWC_2021}
Q.~{Hu}, Y.~{Cai}, Q.~{Shi}, K.~{Xu}, G.~{Yu}, and Z.~{Ding}, ``Iterative
  algorithm induced deep-unfolding neural networks: Precoding design for
  multiuser {MIMO} systems,'' \emph{IEEE Trans. Wireless Commun.}, vol.~20,
  no.~2, pp. 1394--1410, Feb. 2021.

\bibitem{WMMSE2011Shi}
Q.~{Shi}, M.~{Razaviyayn}, Z.~{Luo} \emph{et~al.}, ``An iteratively weighted
  {MMSE} approach to distributed sum-utility maximization for a {MIMO}
  interfering broadcast channel,'' \emph{IEEE Trans. Signal Process.}, vol.~59,
  no.~9, pp. 4331--4340, Sept. 2011.

\bibitem{GJ-MLSP}
J.~Guo and C.~Yang, ``A size-generalizable graph neural network for learning
  multi-user multi-stream {MIMO} precoding,'' \emph{IEEE MLSP}, 2024.

\bibitem{GJ-RGNN}
------, ``Recursive {GNNs} for learning precoding policies with
  size-generalizability,'' \emph{IEEE Trans. Mach. Learn Commun. Netw.},
  vol.~2, pp. 1558--1579, 2024.

\bibitem{ris-dl}
W.~Jin, J.~Zhang, C.-K. Wen \emph{et~al.}, ``Low-complexity joint beamforming
  for {RIS}-assisted {MU-MISO} systems based on model-driven deep learning,''
  \emph{IEEE Trans. Wireless Commun.}, vol.~23, no.~7, pp. 6968--6982, July
  2024.

\bibitem{guo2025attention}
J.~Guo and C.~Yang, ``When attention is beneficial for learning wireless
  resource allocation efficiently?'' \emph{arXiv:2507.02427}, 2025.

\bibitem{RIS_BCD}
H.~Guo, Y.-C. Liang, J.~Chen \emph{et~al.}, ``Weighted sum-rate maximization
  for reconfigurable intelligent surface aided wireless networks,'' \emph{IEEE
  Trans. Wireless Commun.}, vol.~19, no.~5, pp. 3064--3076, May 2020.

\bibitem{LSJ}
S.~Liu, J.~Guo, and C.~Yang, ``Multidimensional graph neural networks for
  wireless communications,'' \emph{IEEE Trans. Wireless Commun.}, vol.~23,
  no.~4, pp. 3057--3073, April 2024.

\end{thebibliography}
	\end{spacing}
	
\end{document}